\documentclass{aa}
\usepackage{graphics}

\begin{document}

\title{Analytic description of the r-mode instability in uniform 
density stars}

\author{K.~D. Kokkotas\inst{1}
\and N. Stergioulas\inst{2}}

\institute{ Department of Physics, Aristotle University of Thessaloniki,
Thessaloniki 54006, Greece \\ e-mail: kokkotas@astro.auth.gr
\and 
Max Planck Institute for Gravitational Physics,
The Albert Einstein Institute,
D-14473 Potsdam, Germany \\ e-mail: niksterg@aei-potsdam.mpg.de} 

\date{Received / Accepted}
\thesaurus{08(08.14.1, 08.15.1, 02.09.1, 02.07.2)}

\titlerunning{r-mode instability in uniform density stars}
\maketitle

\begin{abstract}
  
  We present an analytic description of the $r$-mode instability in
  newly-born neutron stars, using the approximation of uniform
  density. Our computation is consistently accurate to second order in
  the angular velocity of the star.  We obtain formulae for the
  growth-time of the instability due to gravitational-wave emission,
  for both current and mass multipole radiation and for the damping
  timescale, due to viscosity. The $l=m=2$ current-multipole radiation
  dominates the timescale of the instability. We estimate the deviation
  of the second order accurate results from the lowest order approximation
  and show that the uncertainty in the equation of state has only a small
  effect on the onset of the $r$-mode instability. The viscosity
  coefficients and the cooling process in newly-born neutron stars
  are, at present, uncertain and our analytic formaulae enables a
  quick check of such effects on the development of the instability.


\end{abstract}

\section{Introduction}

The recently discovered $r$-mode instability (\cite{nanew})
in rotating neutron stars, has significant implications on the
rotational evolution of a newly-born neutron star.
The $r$-modes are unstable due to the Chandrasekhar-Friedman-Schutz 
(CFS) mechanism (\cite{chandra}, Friedman \& Schutz 1978) 
(see also Friedman \& Morsink 1997).  
Andersson, Kokkotas \& Schutz (1998) (see also Lindblom et al. 1998)  
find that the $r$-mode instability 
is responsible for slowing down a rapidly rotating, newly-born
neutron star to rotation rates comparable to that of the
initial period of the Crab pulsar ($\sim$19 ms) or the recently
discovered 16 ms X-ray pulsar in the supernova remnant
N157B (Marshall et al. 1998) (with an estimated initial period
of 6-9 ms).  
This is achieved by the emission of current-quadrupole gravitational 
waves, which reduce the angular momentum of the star. 
Additionally, as the initially rapidly rotating star spins down, an
energy equivalent to roughly 1\% of a solar mass is radiated in
gravitational waves, which makes the process an interesting source
of detectable gravitational waves (Owen et al. 1998).
 
In the present paper, we investigate the $r$-mode instability to 2nd
order accuracy in the angular velocity of the star, in the
approximation of uniform density (the actual density profile of
realistic neutron stars is nearly uniform), where all results can be
obtained analytically. 
\footnote{The computation in \cite{AKS98} is
carried out in the 2nd order accurate, slow-rotation formalism for
Newtonian stars by Saio (\cite{saio}) while 
\cite{LOM98}, arrive at the same conclusion,
using a first-order, approximate, calculation.} 
While our analytic results provide an
independent check of the numerical results in \cite{AKS98} and
Lindblom et al. 1998, our main objective is to present a simple set of
equations, which enable one to obtain a qualitative insight into the
mechanism of the $r$-mode instability and to quickly check the
dependence of the instability on various important factors, such as
the central density of the star, the different types of viscosity in
neutron stars, the different possible cooling processes etc. 
Additionally, we expect that there is a number of issues related to the
spinning mechanisms of pulsar such as accretion disc induced spin up,
or the creation of millisecond pulsars due to accretion-induced 
collapse of a white dwarf (\cite{AKS98}), for which one can use
the simple relations provided here for a fast but still accurate
evaluation of the various evolution scenarii.

\section{The r-mode Instability}

Oscillations of stars are commonly described by the Lagrangian
displacement vector $\vec \xi$, which describes the displacement of a
given fluid element due to the oscillation. Since $\vec \xi$ is a
vector on the $(\theta,\phi)$ 2-sphere, it can be analyzed into a sum
of spheroidal and toroidal components (or polar and axial components,
in a different terminology).  In a non-rotating star, the usual $f$,
$p$ and $g$ modes of oscillation are purely spheroidal, characterized
by the indeces $(l,m)$ of the spherical harmonic function $Y_l^m$. In
a rotating star, modes that reduce to purely spheroidal modes in the
non-rotating limit, also acquire toroidal components.  Conversely,
$r$-modes in a non-rotating star are purely toroidal modes with
vanishing frequency.  In a rotating star, the displacement vector
acquires spheroidal components and the frequency in the rotating
frame, to first order in the rotational frequency $\Omega$ of the
star, becomes
\begin{equation}
\omega_r = {{2 m \Omega}\over {l (l+1)}} \ ,
 \label{omega_r}
 \end{equation}
for a given ($l,m$) mode. 
 An inertial observer, measures a frequency of 
\begin{equation}
\omega_i = \omega_r - m\Omega \ .
\label{omega_i}
\end{equation}
From (\ref{omega_r}) and (\ref{omega_i}), it can be deduced that, a
counter-rotating (with respect to the star, as defined in the
co-rotating frame) $r$-mode, appears as co-rotating with the star to a
distant inertial observer.  Thus, to O($\Omega$), all $r$-modes with
$l \geq 2$ are generically unstable to the emission of
gravitational radiation, due to the Chandrasekhar-Friedman-Schutz
(CFS) mechanism (note that the $l =1$ $r$-mode is marginally
unstable, to this order). The instability is active for as long as its
growth-time is shorter than the damping-time due to the viscosity of
neutron-star matter.  Its effect is to slow-down, within a year, a
rapidly rotating neutron star to slow rotation rates and this explains
why only slowly-rotating pulsars are associated with supernova
remnants (\cite{AKS98}). Thus, the $r$-mode instability does not allow
millisecond pulsars to be formed after an accretion-induced collapse
of a white dwarf.  It seems that millisecond pulsars can only be
formed by the accretion-induced spin-up of old, cold, neutron stars.

\section{The 2nd-order Accurate Slow Rotation Formalism}

To O($\Omega$), the star is still spherical, and one can only
determine the angular dependence of the $r$-mode eigenfunctions and
their lowest order $r$-dependence (the latter is obtained by taking
the curl of the perturbed equations of motion).  For obtaining the
second-order correction to the eigenfunctions and to the frequency,
one must proceed to a consistent O($\Omega^2$) calculation. We follow the
formalism for computing $r$-modes in Newtonian stars, due to
\cite{saio}, that was presented in more detail in \cite{AKS98}. Here we
will only summarize the equations needed for the uniform density case.
\subsection{Assumptions}

We make the following assumptions: 
\begin{enumerate}
\item the perturbations are adiabatic,
\item the star has uniform density,
\item the rotation of the star is uniform, and
\item  the perturbation of the gravitational potential can be neglected 
     (Cowling approximation), 
\end{enumerate}
These assumptions are justified by the fact that, even for
temperatures $T=10^9K$, the thermal energy of the star is much less
than the Fermi energy of its interior ($> 60$ MeV).  Also, at such
temperatures, the initially differentially rotating proto-neutron star
is rotating uniformly, due to the formation of a solid crust (see
\cite{S98}, for a recent review on rotating neutron stars). The Cowling
approximation has been shown to yield sufficiently accurate results
for $r$-modes in slowly rotating, Newtonian stars (\cite{saio,PBR81}).

\subsection{Definitions}

In a slowly rotating star, the dominant correction to its structure
is of O($\Omega^2$). The analysis of perturbations of the
star is simplified by introducing a new radial coordinate $a$, defined through
\begin{equation}
r = a(1+\epsilon) \ ,
\end{equation}
where $\epsilon=\epsilon(a,\theta)$ is a quantity of O($\Omega^2$),
representing the deformation of the equilibrium structure from the
non-rotating configuration.  In the new coordinate system, all equilibrium 
quantities are functions of $a$ only and the surfaces of
constant $a$ are equipotential surfaces.

Since equilibrium neutron stars are stationary and axisymmetric, a general
oscillation can be analyzed into a sum of normal modes, with harmonic
time-dependence $e^{i(m\phi+\omega_i t)}. $The displacement vector 
for a given $r$-mode can be written as
(\cite{saio}):
\begin{eqnarray}
{\vec \xi}/a = {\bf T} + {\bf S}
         &=& \left(0,K_{l m} \sin^{-1}\theta \partial_{\phi},
          -K_{l m}\partial_\theta \right)Y_l^m \nonumber \\
         &+& \sum_{\nu \mu}
           \left(S_{\nu\mu},H_{\nu\mu}\partial_\theta,
           H_{\nu \mu} \sin^{-1}\theta \partial_{\phi}
          \right)Y_\nu^\mu,
\label{rdisp}
\end{eqnarray}
where ${\bf T}$ and ${\bf S}$ are the {\em toroidal} and {\em
  spheroidal} parts of the displacement, respectively.  Note
that the toroidal part has vanishing $a$-component and is
described only by the function $K_{lm}= K_{lm}(a)$, which multiplies
a toroidal angular vector. The spheroidal part has an non-vanishing 
$a$-component of O($\Omega^2$), described by the functions
$S_{\nu \mu}=S_{\nu \mu}(a)$. The $\theta$ and $\phi$-components
of the spheroidal part are described by the functions $H_{\nu \mu}=H_{\nu \mu}(a)$
(also of O($\Omega^2$)), multiplying spheroidal angular vectors.

The perturbation
in the pressure is expressed in terms of spheroidal radial functions
$\zeta_{\nu \mu}$, as
\begin{equation}
\delta p = \rho g a \sum_{\nu \mu} \zeta_{\nu \mu}Y_\nu^\mu \ ,
\end{equation}
where $\delta p$ is the Eulerian variation in the pressure 
(the variation of the pressure at a fixed point in space), $\rho$ is
the density and $g= -\rho^{-1} dP/d a$ is the acceleration of gravity.

\subsection{The Propensity Rule} 

For zero-temperature (barotropic) stars, it can easily be shown from the
perturbation equations, that only modes with $l=m$ exist. Then,
only the $\nu=l+1$, $\mu=m$ terms contribute in the expansions for
the displacement vector and the perturbation in the pressure (in the
remainder of the text, we will drop the index $m$ in these
quantities).  The absence of $l-1$ terms (the spherical harmonics
$Y_{l-1}^m$ are zero) means that rotation excites only higher
multipole spheroidal parts.  This is in agreement with the
``propensity'' rule suggested by \cite{CF91}, for the oscillations of
slowly rotating relativistic stars, i.e. that the rotational coupling
of a toroidal $l$-term with a spheroidal $l+1$-terms is strongly
favored over the coupling with a spheroidal $l-1$-term. We find that,
in uniform
density stars, the ``propensity'' rule completely eliminates the coupling
to lower-multipole terms.

\subsection{The Perturbation Equations}

A normal-mode solution to the perturbation equations satisfies the
perturbed Euler equations, the perturbed continuity equation and the
relation between the perturbations in density and pressure.
We define dimensionless frequencies as 
\begin{equation}
   \bar \omega_r = \omega_r \Bigl( \frac{R^3}{GM} \Bigr)^{1/2}, 
\end{equation}
and 
\begin{equation}
   \varpi = \Omega \Bigl( \frac{R^3}{GM} \Bigr)^{1/2}, 
\end{equation}
and expand the frequency in the rotating frame as
\begin{equation}
\bar \omega_r = \sigma_0 \varpi + \sigma_2 \varpi^3.  
\end{equation}
Writing the distortion parameter $\epsilon$ as
\begin{equation}
\epsilon \equiv  \left[ \tilde D_1(a) + \tilde D_2(a) P_2(\cos \theta) 
\right] \varpi^2,
\end{equation}
(where $P_2(\cos \theta)$ is the Legendre polynomial) and expanding 
the perturbations
equations consistently to second order in the angular velocity 
of the star, we find that the eigenfunctions 
$\zeta_{l+1}$, $S_{l+1}$, $H_{l+1}$ 
and $K_l$ 
are given by the following set of equations:
\begin{equation}
a{{d\zeta_{l+1}}\over {da}} =   \left(l-1\right)\zeta_{l+1} \ ,
\label{zpert2}\end{equation}
\begin{equation}
  a {{d S_{l+1}}\over {da}} =
  -\left(4 +l\right) S_{l+1} -h\zeta_{l+1} \ ,
\label{spert2}
\end{equation}
\begin{equation}
  H_{l+1} = S_{l+1} + \frac{(2l+1)(l+1)}{8l\sigma_0}
\left[(l+1)\sigma_2
        +6 \tilde D_2 \right] \zeta_{l+1},
\end{equation}
and
\begin{equation}
K_l  = i \frac{(l+1) \sqrt{2l+3}}{2l \bar \omega_r \varpi} \zeta_{l+1}.
\label{K}
\end{equation}
In (\ref{spert2}), $h$ is 
\begin{eqnarray}
h &=& \frac{1}{\sigma_0^2} \Biggl \{ \frac{(l+1)}{l} 
\Bigl[ (2l+3)\frac{\sigma_2}{\sigma_0}
    +6(l-1) \tilde D_2 \Bigr] \nonumber \\
  && +3\left( 3 \tilde D_2 + a \frac{d \tilde D_2}{da} \right) \Biggr\}.
\end{eqnarray}
Note that, the perturbation in the pressure is independent of the 
displacement
vector and can be found by analytically integrating (\ref{zpert2}), 
while  the toroidal
function $K_l$ is given algebraically in terms of the perturbation 
in the pressure. 
The spheroidal function $S_{l+1}$ satisfies a differential 
equation that
depends on the perturbation in the pressure and the structure of 
the star and
can not be obtained analytically, but will not be needed for the 
remainder of this paper. 
The spheroidal function $H_l$ is given algebraically in terms of 
$S_{l+1}$ and $\zeta_{l+1}$.

\subsection{Boundary Conditions}

From the leading terms of $S_l$ and $\zeta_l$ near $a=0$, one 
obtains the boundary condition at the center of the star:
\begin{equation}
(2l+3)S_{l+1} + h \zeta_{l+1}=0 \ .
\label{cbc1}\end{equation}
At the surface of the star, the Lagrangian variation of the pressure
vanishes (a fluid element on the surface of the unperturbed
configuration must also be on the surface of the perturbed
configuration):
\begin{equation}
\Delta p = \delta p + \vec \xi \vec \nabla p =0,  \label{Dp}
\end{equation}
or
\begin{equation}
  \zeta_{l+1} = S_{l+1}. \label{bs}
\end{equation}
To $O(\Omega)$, (\ref{Dp}) is satisfied trivially, while to  $O(\Omega^2)$ it yields
the correction to the eigenfrequency to that order.

\section{Eigenfunctions and Eigenfrequencies}

Eq.  (\ref{zpert2})  for $\zeta_{l+1}$ implies a solution of
the form
\begin{equation}
\zeta_{l+1} \sim a^{l-1}.
\end{equation}
Since $\zeta_{l+1}$ is of order $O(\varpi^2)$, we normalize it to
the dimensionless quantity
\begin{equation}
  \zeta_{l+1} = \varpi^2 \Bigl(\frac{a}{R} \Bigr)^{l-1}.
\end{equation}
Then, (\ref{K}) yields  
\begin{equation}
      K_{l} = i \frac{ (l+1) \sqrt{2l+3} }{2l} 
\Bigl( \frac{\Omega}{\omega_r} \Bigr) 
           \Bigl(\frac{a}{R} \Bigr)^{l-1},
\end{equation}
where $R$ is the radius of the star. 
These are the only two eigenfunctions needed
for the remainder of the paper.

An expression for the eigenfrequency of a given mode can be obtained
from constructing an integral relation using the perturbed Euler
equations (see the appendix of \cite{saio}). For uniform density stars
\begin{equation}
  \sigma_0 = \frac{2}{l+1}, \ \ \ \ {\rm and} \ \ \ \  \sigma_2 
= \frac{5 l (l+1)^2-8}{(l+1)^4},
\end{equation}
or
\begin{equation}
  \bar \omega_r = \frac{2}{l+1} \varpi 
+ \frac{5 l (l+1)^2-8}{(l+1)^4} \varpi^3,
\end{equation}
(cf. appendix II of \cite{PBR81}).  It should be noted that the
uniform density solution in appendix II of Provost et al. 1981 contradicts an
erroneous conclusion in appendix I. of the same paper. The authors
conclude that there are no $r$-modes to $O(\Omega^2$) in a barotropic
star, by showing that the perturbation equations do not admit
solutions with a $Y_l^m(\cos \theta)$ dependence of $\xi_r$. But 
the lowest-order angular dependence of $\xi_r$ is $Y_{l+1}^m(\cos \theta)$.

\section{Dissipation Time-scales}

\subsection{Energy of mode}

The energy of the mode, measured in the rotating frame, is
\begin{eqnarray}
  E &=&  \frac{1}{2} \int \rho | \dot \xi | ^2 dV,  \nonumber \\
    &=&  \frac{l(l+1)}{2} \omega_r^2 \rho \int_0^R a^4 | K_l |^2 da,
\end{eqnarray}
which gives
\begin{equation}
   E  =  \frac{(l+1)^3}{8l} \rho \Omega^2 R^5.
\end{equation}

\subsection{Dissipation due to Gravitational Waves}
The dissipation of energy due to the emission of gravitational waves 
can be estimated from the standard multipole formula:
\begin{equation}
     \left. \frac{dE}{dt} \right|_{\rm gw} 
            = - \omega_r \sum_l N_l \omega_i^{2l+1} 
                              \bigl(  | \delta D_l^m|^2 
                                     + | \delta J_l^m|^2 \bigr),
\label{dw}
\end{equation}
where
\begin{equation}
      N_l = 4 \pi \frac{(l+1)(l+2)}{l(l-1)[(2l+1)!!]^2}.  
\end{equation}
In (\ref{dw})
\begin{equation}
      \delta D_l^m = \int \delta \rho a^l Y_l^{m*} dV, 
\end{equation}
are the mass multipole moments and
\begin{equation}
\delta J_l^m 
= 2 \sqrt{\frac{l}{l+1}} \int a^l(\rho \delta \vec v +
\delta \rho \vec v) \vec Y_{l}^{mB*} d V
\end{equation}
are the current multipole moments, where $\vec v$ is the velocity of 
the fluid and $\vec Y_{l}^{mB*}$ are the ``magnetic'' 
vector harmonics (see Thorne 1980).

\subsubsection{Mass Multipoles}

The dominant mass multipole moment is $\delta D_{l+1}$. 
In uniform density stars, 
the Lagrangian variation of the density vanishes, $\Delta \rho$ =0. 
From the relation
between Lagrangian and Eulerian perturbations of a scalar quantity, 
it follows that
\begin{equation}
 \delta \rho = - \vec \xi \vec \nabla \rho.
\end{equation}
The derivative of the density across the surface is a Dirac 
delta-function  at $a=R$,
thus
\begin{equation}
  \delta \rho = - a S_{l+1} Y_{l+1}^l  \rho \delta(a-R)
\end{equation}
The mass-multipole moment becomes
\begin{equation}
  \delta D_{l+1} = - \varpi^2 \rho R^{l+4},
\end{equation} 
and, being of O($\Omega^2$), it 
contributes to $dE/dt|_{\rm gw}$ an O($\Omega^{2l+8}$) term.

\subsubsection{Current Multipoles}

The dominant current multipole moment is 
\begin{equation}
      \delta J_l 
= 2 l \omega_r \int_0^R \rho a^{l+3} K_{l} da,
\end{equation}
which is
\begin{equation}
      \delta J_l 
= i \frac{(l+1)}{\sqrt{2l+3}} \rho \Omega R^{l+4}.
\end{equation}
The contribution of he dominant multipole moment to $dE/dt|_{\rm gw}$ 
is an O($\Omega^{2l+4}$) and an O($\Omega^{2l+6}$) term.

\subsubsection{Growth-time}

The growth time due to the emission of gravitational waves is
\begin{equation} 
    t_{\rm gw}  =  - \frac{2E}{dE/dt|_{gw}}. 
\end{equation}
Including both the mass and current multipole contributions and keeping
the frequency to O($\Omega^2$), we obtain
\begin{eqnarray}
 t_{\rm gw} &=& - \frac{c^{2l+3}}{G} \frac{\pi (l+1)^3}{3l} 
\Biggl \{ (\sigma_0 +\sigma_2 
     \varpi^2) \bigl[l-\sigma_0 - \sigma_2 \varpi^2 \bigr]^{2l+1} 
\nonumber \\
&& \times \left[ \frac{(l+1)^2}{2l+3}N_l + \left[l-\sigma_0
     -\sigma_2\varpi^2\right]^2 N_{l+1}\varpi^4 \right] \nonumber \\
 && \times M R^{2l} \Omega^{2l+2} \Biggr\}^{-1}. \label{tgw}
\end{eqnarray}
To lowest order in $\Omega$, (\ref{tgw}) reduces to
\begin{equation}
 t_{\rm gw} = - \frac{c^{2l+3}}{24 G} \frac{[(2l+3)!!]^2}{(2l+3)(l-1)^{2l}} 
    \left( \frac{l+1}{l+2} \right)^{2l+2}  
\frac{\Omega^{-2l-2}}{M R^{2l}}. \label{tgw0}
\end{equation}

\subsection{Dissipation due to Shear Viscosity}

The dissipation of energy because of the shear viscosity of neutron 
star matter is
\begin{equation} 
 \left.   \frac{dE}{dt} \right|_{\rm sv}  = - 2\int \eta \delta\sigma^{ab} 
                           \delta \sigma^*_{ab} dV,
\end{equation}
where
\begin{equation}
\delta \sigma_{ab} = {{i\omega_r}\over 2} 
\left( \nabla_a \xi_b + \nabla_b\xi_a -\frac{2}{3} g_{ab}\nabla_c\xi^c \right),
\end{equation}
(see e.g. \cite{il}) and $\eta$ is the shear viscosity coefficient. 
We obtain
\begin{eqnarray}
  \left.{dE\over dt}\right|_{\rm sv}  
&=& - l (l+1) \omega^2_r \eta  \Biggl[ \int_0^R 
       a^2 | a \partial_a K_l|^2 da \nonumber \\
    && +  (l-1)(l+2) \int_0^R a^2 |K_l|^2 da \Biggr],
\end{eqnarray}
which yields
\begin{equation}
  \left. {dE\over dt}\right|_{\rm sv}  
= - \frac{(l+1)^3(l-1)(2l+3)}{4l} \eta \Omega^2 R^3.
\end{equation}
The damping time due to shear viscosity is  
\begin{equation}
    t_{\rm sv} =  \frac{3}{4 \pi(l-1)(2l+3)} \frac{M}{\eta R}. \label{tsv}
\end{equation}
%

\subsection{Dissipation due to Bulk Viscosity}

In a neutron star, bulk viscosity can arise, because of the departure 
from nuclear reaction
equilibrium, such as beta-equilibrium,
during the compression and expansion of matter caused by an oscillation. 
The energy is dissipated at a rate
 \begin{equation}
  \left. {dE\over dt}\right|_{\rm bv}  = - \int \zeta | \delta \sigma|^2 dV,
\end{equation}
where $\zeta$ is the coefficient of bulk viscosity
\begin{equation}
  \delta \sigma = - i \omega_r \frac{\Delta p}{\Gamma p},
\end{equation}
is the expansion of the fluid and $\Gamma$ is the adiabatic index. 
The last relation follows from baryon  conservation in an adiabatic 
oscillation. 
Strictly speaking, in a uniform density star, $\delta \sigma=0$.
But, we assumed the uniform density approximation only to make
calculations easier. 
For the bulk viscosity we use an approximate timescale, 
that has been derived by \cite{cl} for spheroidal oscillations 
in uniform-density stars.
Since the bulk viscosity arises because of the change in density, 
for toroidal oscillations
we use the spheroidal formula, but with $l$ replaced by $l+1$:
\begin{equation}
  \tau_{\rm bv}= \frac{3(2l+5)}{2 \pi (l+1)^3} \frac{\Gamma^4 M}{\zeta R}.
\end{equation}
For the purpose of estimating the bulk viscosity only, $\Gamma$
is taken to be equal to 5, i.e. correspond to that of a stiff (nearly uniform density), 
$N=0.25$ polytrope.

\section{Critical Angular Velocities}

Below the superfluid transition temperature, which is  $T \sim 10^9$ K, 
the shear viscosity is dominated by electron-etectron scattering and an 
approximate formula for the viscosity coefficient is 
\begin{equation}
   \eta = 6 \times 10^{18} \frac{\rho^2_{15}}{T^2_9} \ {\rm g/cm \ s},
\end{equation}
(\cite{cl}), where the notation $\rho_{15}$ means normalization
of the density to $10^{15}$ gr/cm$^3$ and $T_9$ normalization of
temperature to $10^9$K. Above the superfluid transition temperature,
the shear viscosity coefficient due to neutron-neutron interactions is 
\begin{equation}
   \eta = 2 \times 10^{18} \frac{\rho^{9/4}_{15}}{T^2_9} \ {\rm g/cm \ s},
\end{equation}
(\cite{FI79}). The bulk viscosity will be important in hot, newly-born
neutron stars, but its coefficient is not as certain as the
coefficient for shear viscosity.  Sawyer (1989) estimates the bulk
viscosity in neutron star matter, assuming that the star cools through
the modified URCA process and that it is transparent to neutrinos.
The coefficient he obtains is
\begin{equation}
\zeta = 6 \times 10^{25} \rho_{15}^2 \omega_r^{-2} T_9^6 {\rm g/cms}.
\label{sawyer}
\end{equation}
It has been suggested (\cite{LS95}) that for temperatures larger than 
a few times 
$10^9 K$ the neutrino optical depth is still large and the bulk 
viscosity is thus inactive. 
If the star cools through the direct URCA reaction, the bulk viscosity 
will be much larger
than in (\ref{sawyer}), but again only for temperatures for which the 
star is transparent to neutrinos. 
It becomes apparent that,  depending on the cooling process and on the
neutrino optical depth in a newly-born neutron star, the bulk viscosity 
can almost  completely damp non-axisymmetric instabilities or have 
only a small effect on them. 
A more detailed study is needed and departure from equilibrium of other 
interactions (such as interactions
between quarks at lower temperatures) should also be considered. 
For the time being we 
will use  (\ref{sawyer}) as an conservative average of the large
error bars associated with the bulk viscosity. 

Another dissipation mechanism that can affect the instability is the 
superfluid mutual friction.
Estimates by Mendell (1991) and Lindblom \& Mendell (1995), suggests that
mutual friction could suppress the gravitational - radiation - driven 
instability of $f$-modes, when the temperature of the star is
between  $10^7$K$<T<10^9$K. It is not clear whether mutual friction
will have the same effect for $r$-modes, and a new calculation of this
effect is needed.
  
At each value of the temperature of the star, the critical angular 
velocity above which gravitations
radiation has the shortest time-scale, compared to the viscosity 
time-scales, is obtained by
solving the equation
\begin{equation}
\frac{1}{\tau_{\rm gw}} +   \frac{1}{\tau_{\rm sv}}
+\frac{1}{\tau_{\rm bv}}=0.
\end{equation}
We specialize to a specific neutron star model with radius 
$R=12.47 $km and 
mass $M=1.5 M_\odot$ (same as in \cite{AKS98} and \cite{LOM98}). 
The density of the star is
$\rho=3.4 \times 10^{14} {\rm gr}/{\rm cm}^3$. 

Fig. 1, shows the critical angular velocity as a function of
temperature (in units of the angular velocity at the mass-shedding
limit for Newtonian, uniform density and uniformly rotating stars,
$\Omega_K \simeq 0.67 \sqrt{\pi G \rho})$.  The solid curve
corresponds to the $O(\Omega^2)$ equation (\ref{tgw}).  A
rapidly-rotating neutron star, born at temperatures $~10^{11}$K loses
angular momentum because of the $r$-mode instability and slows-down. The
minimum angular velocity it could reach is of 
$\Omega_{\rm c} = 414 {\rm s}^{-1}$ (or
period of 15ms) at $T \simeq1 \times 1.5 \times 10^9$ K.  
The mass of the neutron star,
or the adiabatic index, do not have a significant effect on the
minimum critical angular velocity.  The radius of the neutron star,
however, (which can range from 10km to 15km), does have a considerable
effect and the radius of our model represents a mean value of the
expected radius of a typical $1.5 M_\odot$ neutron star. 

For low rotation rates, one can use (\ref{tgw0}) to construct approximate 
equations for the two parts of the curve in Fig. 1. 
The part of the critical curve 
where the shear viscosity dominates can be approximated by 
\begin{equation} 
 \Omega_c^{\rm (sv)} = 581 
 \left( \frac{10 {\rm km}}{R} \right)^{3/2}
\left( \frac{10^9 {\rm K}}{T} \right)^{1/3} \ {\rm s^{-1}},
\end{equation}
wihle the bulk viscosity dominated part is described by 
\begin{equation}
\Omega_c ^{\rm (bv)} = 362
 \left( \frac{R}{10 {\rm km}} \right)^{9/8} 
 \left( \frac{T}{10^9 {\rm K}} \right)^{3/4} \ {\rm s^{-1}}.
\end{equation}
The two approximate expressions are shown as dotted curves in Fig. 1. 
For a period $1.56$ ms (the period of the fastest known millisecond pulsar), 
the lowest order critical angular velocity differs from the $O(\Omega^2)$ 
result by $\sim 17 \%$.

In a similar way, the lowest order approximations to the dissipation 
timescales are
\begin{equation}
 t_{\rm gw} = 1.4 \times 10^6    \left( \frac{10^3 s^{-1}}{\Omega} \right)^6
\left( \frac{1.4 M_\odot}{M} \right)
\left( \frac{10 {\rm km}}{R} \right)^4 \ {\rm s},
\end{equation}
\begin{equation}
 t_{\rm sv} = 3.6 \times 10^7    \left( \frac{R}{10 {\rm km}} \right)^5
\left( \frac{T}{10^9 {\rm K}} \right)^2 \left( \frac{1.4 M_\odot}{M} \right)
\ {\rm s},
\end{equation}
and
\begin{eqnarray}
 &&t_{\rm bv} = 4.6 \times 10^9 \nonumber \\
 && \times \left( \frac{R}{10 {\rm km}} \right)^5
\left( \frac{\Omega}{10^3 s^{-1}} \right)^2
\left( \frac{1.4 M_\odot}{M} \right) \left( \frac{10^9 {\rm K}}{T} \right)^6
\ {\rm s},
\end{eqnarray}

Our current results for the onset of the $r$-mode instability
correspond to neutron stars with a very stiff equation of state. The
results in Andersson et al (1998), correspond to a much softer
equation of state (an N=1.0 polytrope) and a comparison is shown in
Fig. 2. The minimum critical temperature is roughly the same for both
equations of state, although it occurs at a somewhat smaller
temperature in the uniform density case. This shows that the
uncertainty in the equation of state does not have a significant
impact on the $r$-mode instability.

\begin{figure}
\resizebox{\hsize}{!}{\includegraphics{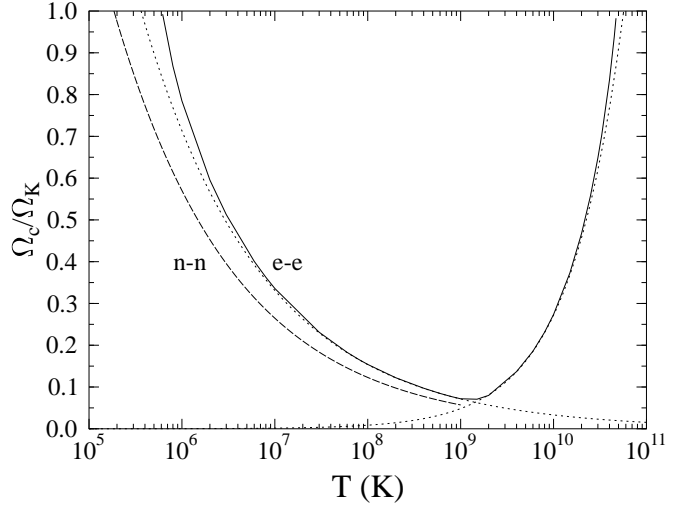}}
\caption{Critical angular velocity for the onset of the $r$-mode 
  instability as a function of temperature (for a 1.5 $M_{\odot}$
  neutron star model).  The solid line corresponds to the
  $O(\Omega^2)$ result using superfluid ($e^- - e^-$) shear viscosity,
  and Sawyer's (1989) estimate for the bulk viscosity. Dotted lines are lowest order
  approximations, while the dashed line corresponds to normal matter
  ($n-n$) shear viscosity.}
\end{figure}

\begin{figure}
\resizebox{\hsize}{!}{\includegraphics{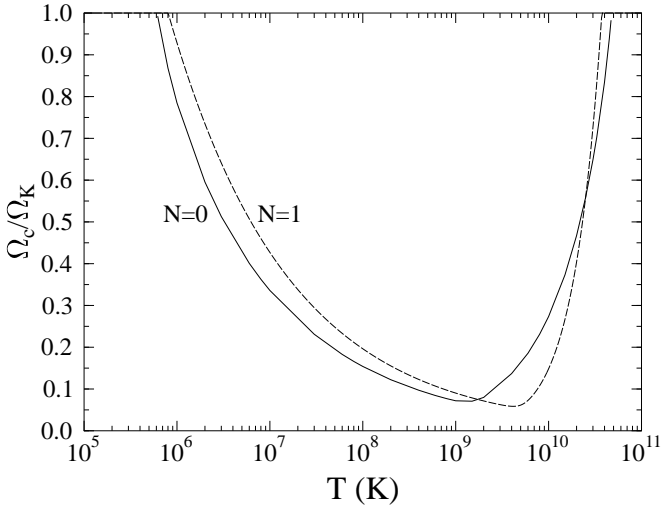}}
\caption{Critical angular velocity for the onset of the $r$-mode 
instability as a function of temperature (for a 1.5 $M_{\odot}$ neutron 
star model). The present  uniform density result ($N=0$) is compared
to the critical curve for the equation of state used in Andersson et al. (1998), 
($N=1.0$ polytrope). The minimum value of $\Omega_c/\Omega_K$ is
roughly the same, and the effect of the equation of state is mainly to shift 
the critical curve to different temperatures.} 
\end{figure}

\section{Discussion}

Our analytical results for the onset of the $r$-mode 
instability in neutron stars agree well with numerically 
obtained results, for the same neutron star model. 
Using our analytic formulae, the uncertainty in the bulk viscosity can 
be easily explored  for different present
and future estimates of the bulk viscosity coefficient. 
The shear viscosity for lower temperatures
is also uncertain, since a high shear viscosity due to the mutual 
friction between superfluid vortices below the superfluid transition 
temperature could suppress the instability.
If future investigations provide a definite answer on the effect of 
mutual friction in a superfluid, one can easily study the implication 
on the $r$-mode instability using the analytic formulae presented 
in this paper.

We would also like to point out that, although an $l=1$ dipole mode 
does not radiate in a non-rotating star, 
it does emit gravitational waves 
through the coupling to higher order terms, in rotating stars. 
According to (\ref{tsv}), the shear viscosity for $l=1$ vanishes 
and thus cannot affect the emission of gravitational waves,
this is not true for realistic equations of state but still the
damping times are extremelly long. 
The $l=1$ $r$-mode will radiate through the coupling to
spheroidal $l=2$ terms, i.e. it will generate mass quadrupole 
radiation. 
The frequency of this mode in the rotating frame 
is \ $\bar \omega_r = \varpi$,  while in the inertial frame 
the frequency is \ $\bar \omega_i = (3/4) \varpi^3$. 
According to the criterion for the onset of the CFS-instability, 
the $l=1$ $r$-modes are thus stable to the emission of 
gravitational waves, 
in contrast to the $l \geq 2$ modes.
Such stable oscillations, unaffected by shear viscosity at low 
temperatures, 
could be excited during a neutron star glitch. 
In analogy to the $l=1$ $r$-modes, an $l=1$ spheroidal
mode, like the $f$ mode or the $p$-modes, will emit current quadrupole 
radiation and this case needs further study.

\begin{acknowledgements}

We would like to thank N. Andersson, J.~L. Friedman and B.~F. Schutz 
for helpful discussions and suggestions.
N.S. acknowledges the generous hospitality of the Max-Planck-Institute 
for Gravitational Physics in Potsdam, Germany.

\end{acknowledgements}


\end{document}